\documentclass[conference]{IEEEtran}
\IEEEoverridecommandlockouts
\usepackage{bm}
\usepackage{graphicx} 
\usepackage{multirow}
\usepackage{algorithm}
\usepackage{algpseudocode}
\usepackage{amsmath}
\usepackage{graphics}
\usepackage{epsfig}
\usepackage{booktabs}
\usepackage{cite}
\usepackage{hyperref}
\usepackage{threeparttable}
\usepackage{xcolor}
\usepackage{colortbl}
\definecolor{mygray}{gray}{0.9}
\definecolor{mypink}{rgb}{.99,.91,.95}
\definecolor{mycyan}{cmyk}{.3,0,0,0}
\def\BibTeX{{\rm B\kern-.05em{\sc i\kern-.025em b}\kern-.08em
		T\kern-.1667em\lower.7ex\hbox{E}\kern-.125emX}}
\usepackage[justification=centering]{caption}
\begin{document}

\title{Reinforced Bit Allocation under Task-Driven Semantic Distortion Metrics\\
	{\footnotesize }
	\thanks{This work was supported in part by NSFC under Grant U1908209, 61571413, 61632001 and the National Key Research and Development Program of China 2018AAA0101400.
	}
}
\author{\IEEEauthorblockN{Jun Shi, Zhibo Chen}
	\IEEEauthorblockA{\textit{CAS Key Laboratory of Technology in Geo-spatial Information Processing and Application System} \\
		\textit{University of Science and Technology of China }\\
		Hefei, China \\
		chenzhibo@ustc.edu.cn}
}

\maketitle
\begin{abstract}
Rapid growing intelligent applications require optimized bit allocation in image/video coding to support specific task-driven scenarios such as detection, classification, segmentation, etc. Some learning-based frameworks have been proposed for this purpose due to their inherent end-to-end optimization mechanisms. However, it is still quite challenging to integrate these task-driven metrics seamlessly into traditional hybrid coding framework. To the best of our knowledge, this paper is the first work trying to solve this challenge based on reinforcement learning (RL) approach. Specifically, we formulate the bit allocation problem as a Markovian Decision Process (MDP) and train RL agents to automatically decide the quantization parameter (QP) of each coding tree unit (CTU) for HEVC intra coding, according to the task-driven semantic distortion metrics. This bit allocation scheme can maximize the semantic level fidelity of the task, such as classification accuracy, while minimizing the bit-rate. We also employ gradient class activation map (Grad-CAM) and Mask R-CNN tools to extract task-related importance maps to help the agents make decisions. Extensive experimental results demonstrate the superior performance of our approach by achieving 43.1\% to 73.2\% bit-rate saving over the anchor of HEVC under the equivalent task-related distortions.

\end{abstract}

\begin{IEEEkeywords}
HEVC, intra coding, bit allocation, reinforcement learning
\end{IEEEkeywords}

\section{Introduction}
In the past few years, continuous progress has been made in the field of image/video analysis and understanding, promoting many intelligent applications such as surveillance video analysis, medical image diagnosis, mobile authentication, etc. This brings out the requirements for efficient compression of image/video signals to support these intelligent applications, which have quite different distortion metrics compared with the traditional image/video compression scenarios.

The development of image/video coding technology has been continuously improving the rate-distortion performance, \textit{i.e.}, reducing the bit-rate under the same quality or reducing the distortion using the same bit-rate. However, the problem is how to define the distortion. Generally, it can be classified into three levels of distortion metrics: \textbf{Pixel Fidelity}, \textbf{Perceptual Fidelity}, and \textbf{Semantic Fidelity}, according to different levels of human cognition on image/video signal \cite{chen2019learning}.
\begin{figure}[]
	\centerline{\includegraphics[scale=0.65]{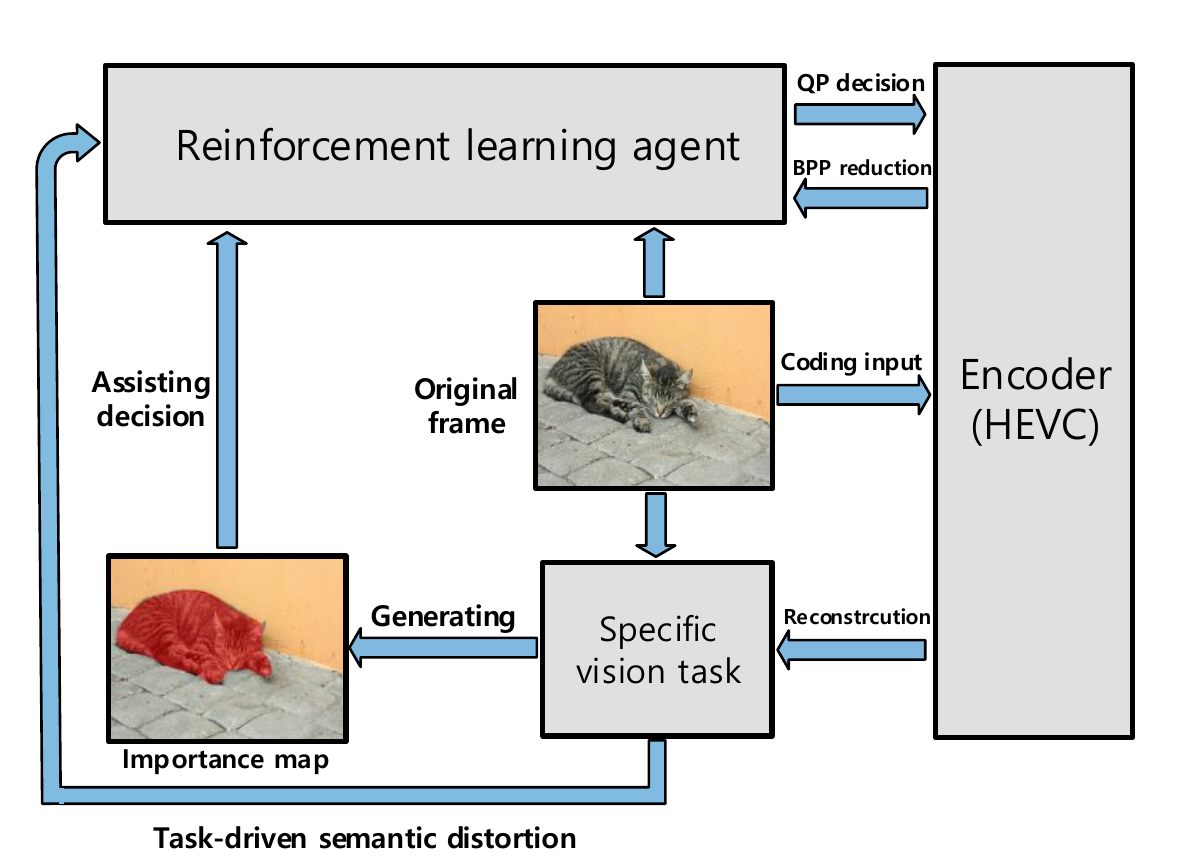}}	
	\caption{Illustration of proposed framework. RL agent decides the QP of each CTU according to tasks.}
	\vspace{-0.4cm}
\end{figure}

\textbf{Pixel Fidelity} metrics like MSE (Mean Square Error) and PSNR (Peak Signal to Noise Ratio) have been widely adopted in many coding techniques and standards (\textit{e.g.}, H.264\cite{1218189}, High Efficiency Video Coding (HEVC)\cite{6316136}, etc.), which can be easily integrated into image/video hybrid compression framework as an in-loop metric for rate-distortion optimization, while it is obvious that pixel fidelity metric cannot fully reflect human perceptual viewing experience \cite{wang2009mean}. Therefore, many researchers have developed \textbf{Perceptual Fidelity} metrics to investigate objective metrics measuring human subjective viewing experience \cite{chen2015hybrid,zhou2019dual}, such as the saliency-guided PSNR\cite{li2011visual}. Based on this new metric, many approaches\cite{prakash2017semantic, hadizadeh2013saliency,khanna2015perceptual,guo2009novel,leung2009perceptual, li2017closed,zhu2018spatiotemporal, oh2012video} optimized the bit allocation scheme to provide better perceptual quality. The common way to achieve this goal in these methods is to allocate relatively more bits in the region-of-interest (ROI) area to ensure acceptable quality in these regions heuristically.

With the development of aforementioned intelligent applications, image/video signals will be captured and processed not only for human eyes, but also for semantic analyses. Consequently, there will be more requirements on research for \textbf{Semantic Fidelity}  metrics to study the semantic difference (\textit{e.g.}, accuracy difference of specific intelligent tasks) between the original image/video and the compressed one.

These various distortion metrics measure the quality of reconstructed image/video content, but it is a contradictory that most of these complicated quality metrics with high performance are not able to be integrated easily into any existing image/video compression frameworks. Some approaches\cite{alakuijala2017guetzli, liu2017recognizable} tried to do this by adjusting image compression parameters (\textit{e.g.}, quantization parameters (QPs)) heuristically according to the embedded quality metrics, but they are still heuristic solutions without ability to automatically and adaptively optimize encoding configurations according to different complicated distortion metrics. \cite{chen2019learning} is the first scheme trying to solve this challenge by designing an end-to-end image coding framework which inherently provides feasibility on integrating complicated metrics into coding loop. But it’s still a huge challenge for traditional hybrid coding frameworks which are not derivable like end-to-end image coding framework.

In this paper, we try to solve this problem by using reinforcement learning (RL) to optimize the bit allocation scheme for HEVC intra coding according to the task-driven distortion metrics, as Fig. 1 shows. We formulate the bit allocation scheme, \textit{i.e.}, deciding the QP of each CTU in sequence, as a Markovian Decision Process (MDP), and then introduce RL to decide the QP of each coding tree unit (CTU) to provide a better bit allocation scheme for the different vision tasks including classification, detection and segmentation. For the specific task, importance maps of the original frames are generated using gradient class activation map (Grad-CAM)\cite{selvaraju2017grad} and Mask R-CNN\cite{he2017mask} tools, which can help the agents make better decisions. In order to train the RL agents efficiently, we establish a universal task-driven bit allocation dataset. Using this dataset, the off-line training can be efficient. With our scheme, we can achieve the bit-rate reduction from 43.1\% to 73.2\% with comparable task-driven distortion, depending on the type of tasks. It shows that the scheme we propose is effective and efficient. 
\section{Task-Driven Bit Allocation Framework}

\subsection{Importance Map Generation}

Although it is possible to train RL agents to directly output the bit allocation scheme merely based on the frame itself, it will make the agents hard to train and lack generalization and extendibility. Therefore, we employ existing approaches to generate the importance maps, which can help and ease the agents to make decisions, as shown in Fig. 2.

First, we implement the Grad-CAM, which uses the gradient backpropagation to flow into the final convolutional layer of the CNN model to produce a localization map highlighting the important regions for predicting the concept. The specific CNN model we adopt is VGG-16\cite{simonyan2014very}, thus we can obtain the importance maps of the frames for the classification task.

Then, we employ Mask R-CNN to get the detection and segmentation results of the original frames. This can help the agents to discriminate the foreground and the background, and further get the density distribution of the instances. These importance maps can be used for the detection and segmentation tasks.
\begin{figure}[!h]
	\vspace{-0.3cm}
	\centerline{\includegraphics[scale=0.4]{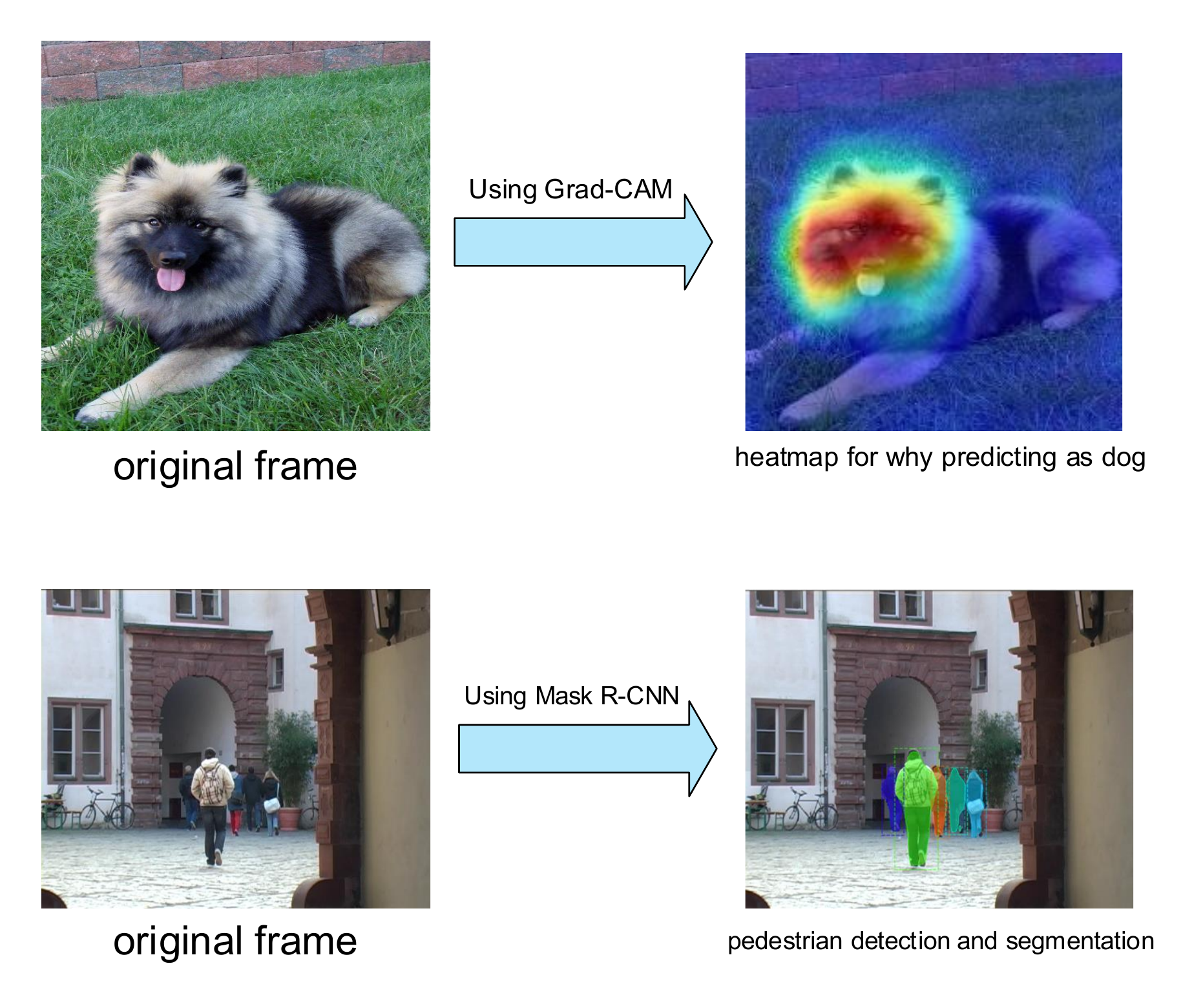}}	
	\caption{Importance map generation.}
	\vspace{-0.2cm}
\end{figure}
Based on this pre-processed information, RL agents are able to  make QP decision precisely and accurately.

\subsection{Reinforcement Learning for Bit Allocation}
After getting the importance maps of the frames, a simple method to optimize towards the task-driven distortion metrics is to increase the bit-rate for the highly weighted area heuristically, such as the threshold scheme. However, these heuristic methods can hardly get the optimal results and may introduce limitation due to the fact that the results rely heavily on the handcrafted designs. Recently, RL has achieved outstanding performance in many tasks, especially in the unsupervised or semi-supervised scenarios. It has also been used in many approaches\cite{hu2018reinforcement,li2019reinforcement,costero2019mamut, iranfar2018machine} to optimize the traditional hybrid coding framework. In this paper, we adopt the RL algorithm, Deep Q-learning (DQN)\cite{mnih2013playing}, to solve this problem.

\begin{figure*}[h]
	\centerline{\includegraphics[scale=0.55]{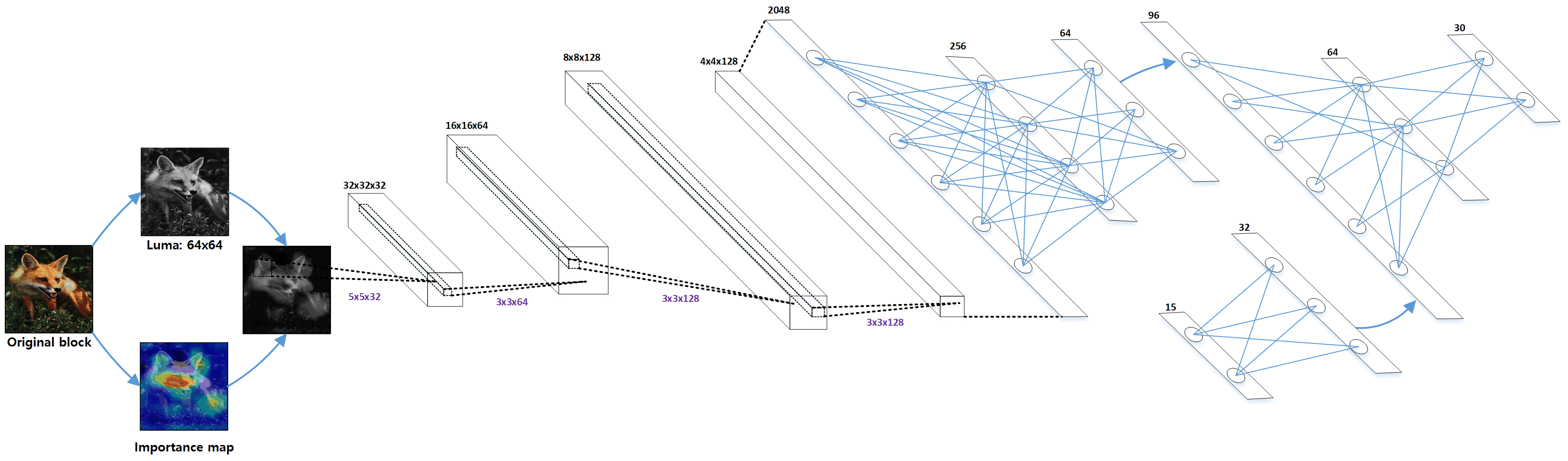}}	
	\caption{Structure of the proposed Q-network. There are two input branches: current CTU part and global information part. The luminance and importance map of current block are concatenated in depth to represent the local information.}
\end{figure*}
\subsubsection{Problem formulation}
We formulate bit allocation problem as a MDP, which includes four elements: \textit{state}, \textit{action}, \textit{reward} and \textit{policy (agent)}. In this process, DQN trains an \textit{agent} that observes the \textit{states} from the environment, and then execute a series of \textit{actions} (QP decisions) in order to optimize the goal. Generally, the goal can be described by the maximization of expected cumulative \textit{reward}. In bit allocation, the specific goal is, given a frame that consists of $i$ CTUs, to generate $i$ QPs for each CTU so that we can achieve the largest bit-rate reduction under the least task-driven distortion. All elements of the MDP are detailed below.

\subsubsection{State}
The agent needs to observe the CTU and then make the QP decision. The CTU information are sent to the agent according to the encoding sequence, from left to right and top to bottom. In this paper, the luminance and the importance map of current CTU are the part of the state. We also include a 15-d feature vector in the state to reflect the global information. The detail of the feature vector is shown in TABLE I.

\begin{table}[h]
	\renewcommand{\arraystretch}{1.5}
	\centering
	\small
	\caption{Global Information Vector}
	\begin{tabular}{l|ll}
		\cline{1-2}
		index & vector components                   &  \\ \cline{1-2}
		1     & number of overll CTUs                &  \\ \cline{1-2}
		2     & index of current CTU                &  \\ \cline{1-2}
		3     & mask ratio of current CTU           &  \\ \cline{1-2}
		4-7   & mask ratio of neighboring CTUs      &  \\ \cline{1-2}
		8     & mask ratio of overall frame         &  \\ \cline{1-2}
		9     & instance number of current CTU      &  \\ \cline{1-2}
		10-13 & instance number of neighboring CTUs &  \\ \cline{1-2}
		14-15 & QPs of left and above CTUs          &  \\ \cline{1-2}
	\end{tabular}
\vspace{-0.2cm}
\end{table}
\subsubsection{Action}
The original QP of HEVC ranges from 1 to 51. Lower QP means more bit-rate,  and more bit-rate means less distortion. However, we do not need all QPs in this paper. When the bit-rate is high enough, there is almost no semantic distortion, which means more bit-rate is useless, although it can still improve the pixel level performance, such as PSNR. 

In our experiment, we observe that the semantic distortion becomes really negligible compared with the original uncompressed images after QP reaches 22. Therefore, we set the action space as QP 22 to 51 in this paper.

\subsubsection{Reward}
The cumulative reward is the optimization goal of RL agent. Like the rate-distortion cost in HEVC coding, the designed reward has a similar format:
\begin{equation}
reward=\lambda\cdot BPP_{save} - Distortion_{task}.
\end{equation}
where $BPP_{save}$ means the reduction of bit per pixel (BPP) in current CTU from the anchor of QP = 22 to the decision QP. The $Distortion_{task}$ is a penalty term for the semantic difference of the task result, such as the detection number difference for the detection task, and $\lambda$ is a Lagrange factor to balance the BPP reduction and the risk of semantic loss.

\subsubsection{Agent}
The agent is a Q-network, which is used for optimal QP prediction. Taking the state $s_t$ as input, the Q-network will output the decayed cumulative reward (Q-value) of each action $a$ as $Q(s_t, a)$. We can get the optimal action $a_t^*$ as :
\begin{equation}
a_t^* = \mathop{\arg\max}_{a} Q(s_t, a).
\end{equation}

For the Q-network structure, we have two input branches: the current CTU part and the global feature vector. The CTU information flows through four convolutional layers to extract the features. Then we concatenate the features with the global feature vector after ascending dimension together. The combination of these features can help better understand the content. Next, the overall features will flow through three fully-connected layers, including two hidden layers and one output layer. All convolutional layers and hidden fully-connected layers are activated with Leaky Rectified Linear Unit (LeakyReLU), while the output layer is not activated. Fig. 3 shows the detail of the proposed Q-network.

\begin{figure*}[!h]
	\centering{\includegraphics[scale=0.185]{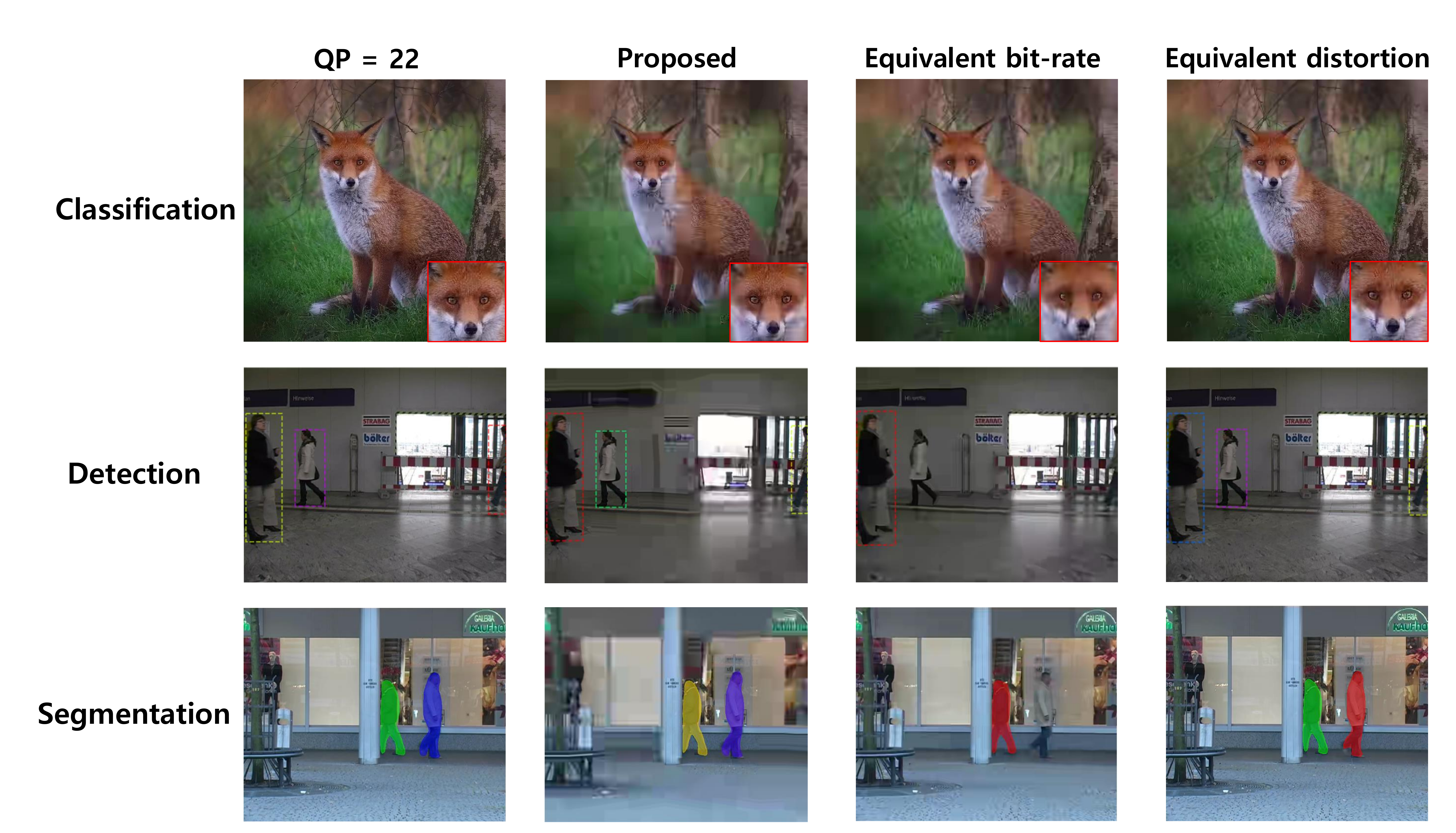}}	
	\caption{Detailed comparison between the proposed method and the original HEVC. The first, third and fourth columns are encoded with fixed QP, while the second column is the result using our task-driven bit allocation.}
	\vspace{-0.2cm}
\end{figure*}

\section{Experimental Results and Analysis}
In this section, we will introduce the dataset and the specific experimental details for task-driven bit allocation. 

\subsection{Task-driven Bit Allocation (TBA) Dataset}
As we employ RL agents (CNN models) as the bit allocation predictors, we need a large amount of training data. So we establish a universal dataset for Task-driven Bit Allocation, namely TBA dataset. First, for detection and segmentation, we collect images from TUD-Brussels pedestrian dataset\cite{wojek2009multi}, including 1000 images. And for classification, we select 2400 high resolution images (larger than 576$\times$576, this is for better bit allocation since it is on CTU level) from the ImageNet\cite{deng2009imagenet} test set, and then resize them to 576$\times$576. Then all the images are encoded by the HEVC reference software HM 16.9\cite{HM} with all-intra main configuration, while all QPs in 22 $\sim$ 51 are applied for encoding. During encoding, we collect the BPP and MSE of every CTU for further usage.

After encoding, we get the reconstruction of all images, which are then sent to the vision task CNNs to get the semantic results. We use VGG-16 for classification and Mask R-CNN for detection and segmentation.  Finally, the TBA dataset is obtained, which is randomly divided into training (80\%) and test (20\%) sets.

\subsection{Configuration of Experiments}

In this part,  we implement our approach into the HEVC reference software HM 16.9. In HM, the all-intra main configuration was used. For comparison, we run the fixed QP = 22 as benchmark, since the semantic distortion under this QP is really negligible. In addition, two experiments of fixed QP scheme without bit allocation for the equivalent bit-rate and equivalent distortion with the proposed approach are also done.

 After coding, we sent the reconstructions to the task CNNs to measure the semantic task-driven distortions. This distortion is defined as the semantic similarity, \textit{i.e.}, the difference of outputs compared with the benchmark frames of  QP = 22. For classification in ImageNet, the top-5 accuracy is adopted, while the number of accurate detected instances is used for detection task. And the intersection over union (IOU) is measured for the  segmentation task. Fig. 4 and TABLE II show the results of our experiments.


\subsection{Performance Evaluation and Analysis}

\textbf{Assessment of bit-rate reduction.}
It is interesting to investigate how many bits can be saved when applying our approach compared with the original HEVC.  Taking QP = 22 as the benchmark, we compare the BPP of our scheme with the one which possesses the similar semantic distortion (Baseline* in TABLE II). For classification task, we can further save 43.1\% bits (relative ratio) over this fixed QP strategy under the same distortion. Similarly, our approach saves 73.2\% bits for the pedestrian detection, while for segmentation, more bits should be used to preserve the pixels of foreground, especially the boundary, so only 58.6\% bits are saved, less than the result of detection task. The RL agent is very astute to use as few bits as possible while minimizing the task-driven distortion. 

\textbf{Evaluation of task-driven distortion.}
We also investigate how our approach can preserve the semantic fidelity to minimize the task-driven distortion using the same bits. We choose the fixed QP of HEVC to encode the frames using equivalent BPP (Baseline in TABLE II) to examine the effectiveness of our proposed scheme. We can see that our bit allocation strategy leads to much less semantic distortions in all tasks. For classification,  our approach uses more bits on the important regions, such as the face of fox in Fig. 4. Consequently, the task-driven distortion  is much smaller. Similarly, for detection and segmentation, our approach can better preserve the quality of foregrounds. Obviously, our RL agent can spend the bits rationally according to the tasks.

\begin{table}[]
	\renewcommand{\arraystretch}{1.5}
	\centering
	\footnotesize
	\caption{Bit-Rate Reduction Ratio and Task-Driven Distortion Compared with QP = 22}
	\begin{tabular}{l|cc|cc|cc}
		\hline
		& \multicolumn{2}{c|}{Proposed}                                                                                & \multicolumn{2}{c|}{Baseline}                                                                           & \multicolumn{2}{c}{Baseline*}                                                                           \\ \hline
		Vision task    & \begin{tabular}[c]{@{}c@{}}BR\\ \end{tabular} & \begin{tabular}[c]{@{}c@{}}DIST\\ \end{tabular} & \begin{tabular}[c]{@{}c@{}}BR\\ \end{tabular} & \begin{tabular}[c]{@{}c@{}}DIST\\ \end{tabular} & \begin{tabular}[c]{@{}c@{}}BR\\ \end{tabular} & \begin{tabular}[c]{@{}c@{}}DIST\\ \end{tabular} \\ \hline
		Classification & 85.2\%                                             & 3.7\%                                                & 87.2\%                                             & 12.2\%                                               & 74.0\%                                             & 3.5\%                                                \\ \hline
		Detection      & 80.2\%                                             & 2.7\%                                                & 79.3\%                                            & 18.6\%                                               & 26.2\%                                             & 2.2\%                                                \\ \hline
		Segmentation   & 66.2\%                                             & 6.8\%                                                & 66.4\%                                             & 11.9\%                                               & 18.5\%                                             & 6.8\%                                                \\ \hline
	\end{tabular}
	\begin{tablenotes}
		\item Baseline and Baseline* indicate the equivalent bit-rate and distortion schemes, resepectively. Bit-rate reduction ratio and semantic distortion are abbreviated as BR and DIST in this table.
	\end{tablenotes}
	\vspace{-0.4cm}
\end{table}

\section{Conclusion}
In this paper, we propose an automatically optimized task-driven bit allocation scheme for HEVC intra coding using reinforcement learning. We formulate sequential QP decision for each CTU as a Markovian Decision Process and then train RL agent to integrate the vision task-driven distortion metrics into HEVC. Benefiting from the Grad-CAM and Mask R-CNN tools, we obtain the task-related importance maps of the original frames before bit allocation. These importance maps can do help for the RL agents. In addition, we establish a TBA dataset, with which, off-line training can be efficient. Compared with the original HEVC, our approach can save about 43\% to 73\% bits under the equivalent task-driven distortion. In future work, we will consider extending our scheme to inter mode for further  optimization.

\bibliographystyle{IEEEtran}
\bibliography{IEEEexample,reference}





\end{document}